\documentclass[10pt,conference]{IEEEtran}

\usepackage{cite}
\usepackage{amsmath,amssymb,amsfonts}
\usepackage{algorithm}
\usepackage{algorithmic}
\usepackage{graphicx}
\usepackage{booktabs}
\usepackage{array}
\usepackage{multirow}
\usepackage{textcomp}
\usepackage{xcolor}
\usepackage[hyphens]{url}
\usepackage[hidelinks]{hyperref}
\usepackage{xspace}
\hypersetup{
  pdfauthor={Runhao Liu, Minnan Pei, Fei Ding, Guangzhen Yao, You Li, Peng Xiao, Gang Li, Peng Zhang},
  pdftitle={LATTICE: Constraint-Directed Scheduling, Memory Planning, and Pipeline Refinement for NPUs},
  pdfsubject={},
  pdfkeywords={NPU compilation, command scheduling, static memory planning},
  pdfcreator={}
}

\title{LATTICE: Constraint-Directed Scheduling, Memory Planning, and
Pipeline Refinement for NPUs}
\author{
\IEEEauthorblockN{
Runhao Liu\IEEEauthorrefmark{1}\IEEEauthorrefmark{6},
Minnan Pei\IEEEauthorrefmark{2}\IEEEauthorrefmark{6},
Fei Ding\IEEEauthorrefmark{3},
Guangzhen Yao\IEEEauthorrefmark{4},
You Li\IEEEauthorrefmark{5},
Peng Xiao\IEEEauthorrefmark{1},
Gang Li\IEEEauthorrefmark{2}, and
Peng Zhang\IEEEauthorrefmark{1}}
\IEEEauthorblockA{
\IEEEauthorrefmark{1}Zhejiang University \quad
\IEEEauthorrefmark{2}Institute of Automation, Chinese Academy of Sciences\\
\IEEEauthorrefmark{3}Alibaba Group \quad
\IEEEauthorrefmark{4}Northeast Normal University \quad
\IEEEauthorrefmark{5}Guangdong Polytechnic Normal University\\
{\footnotesize runhaoliu@zju.edu.cn, peiminnan19@mails.ucas.ac.cn}}
}

\newcommand{\lattice}{\textsc{Lattice}\xspace}
\newcommand{\mpas}{\textsc{Mpas}\xspace}
\newcommand{\dlr}{\textsc{Dlr}\xspace}
\newcommand{\cpe}{\textsc{Cpe}\xspace}
\newcommand{\notice}[1]{#1}

\begin{document}
\maketitle
\pagestyle{plain}
\thispagestyle{plain}
\begingroup
\renewcommand{\thefootnote}{\fnsymbol{footnote}}
\footnotetext[6]{Runhao Liu and Minnan Pei contributed equally to this work.}
\endgroup

\newcommand{\OverallPeakVsReference}{18.3\% lower}
\newcommand{\OverallDDRVsReference}{20.4\% lower}
\newcommand{\OverallSpillVsReference}{14.1\% lower}
\newcommand{\OverallMakespanVsReference}{16.3\% lower}
\newcommand{\OverallPeakVsBase}{38.3\% lower}
\newcommand{\OverallDDRVsBase}{62.0\% lower}
\newcommand{\OverallSpillVsBase}{64.9\% lower}
\newcommand{\OverallMakespanVsBase}{57.5\% lower}
\newcommand{\OverallBestOrTiedCells}{24}
\newcommand{\OrderLogicalRange}{21.2\%--81.2\%}
\newcommand{\OrderPhysicalRange}{1.0\%--77.2\%}
\newcommand{\OrderDDRRange}{49.5\%--67.7\%}
\newcommand{\OrderSpillRange}{44.1\%--60.8\%}
\newcommand{\OrderMakespanRange}{45.6\%--61.7\%}
\newcommand{\MPASLogicalStageChange}{53.4\% lower}
\newcommand{\DLRDDRStageChange}{6.6\% lower}
\newcommand{\DLRSpillStageChange}{4.4\% lower}
\newcommand{\CPEGeomeanReduction}{12.1\%}
\newcommand{\CPEImprovedWorkloads}{6}
\newcommand{\IgnoreInvalidWorkloads}{6}
\newcommand{\IgnoreViolationTotal}{36,399}
\newcommand{\ReuseEdgeTotal}{550,492}
\newcommand{\PrecedenceEdgeTotal}{165,562}
\newcommand{\ReuseToPrecedenceRatio}{3.3$\times$}
\newcommand{\FreezeMedianOverIgnore}{93.9\%}
\newcommand{\PostPlanStaticErrorTotal}{265,988}
\newcommand{\PostPlanInvalidWorkloads}{6}
\newcommand{\CompileTotalRange}{2{,}353.7--434{,}607.5 ms}

\begin{abstract}
General-purpose NPUs execute fine-grained command DAGs across heterogeneous
compute and memory-transfer engines backed by finite, explicitly managed
on-chip memories.  This execution model creates a directed dependency between
scheduling and memory planning: different legal topological orders induce
different lifetime overlap, placement opportunities, and spill behavior,
while a materialized layout introduces physical-address reuse constraints
absent from the input precedence DAG\@.  Command order therefore shapes the
feasible memory plan, and the realized plan in turn defines the legal space
for subsequent timing refinement.

We present \lattice, a deterministic constraint-directed compiler pipeline.
Memory-Pressure-Aware Topological Scheduling reshapes lifetime geometry before
address binding; Deterministic Linear Repackaging materializes tiered
placement, spill/reload events, and plan-induced reuse constraints; and
Critical Path Enhancement recovers pipeline parallelism while preserving the
selected memory plan.  Every accepted schedule passes independent memory and
timing verification.  Across six artifact-provided command traces labeled as
derived from a Da Vinci NPU flow, \lattice achieves the best or tied-best
result in all \textbf{\OverallBestOrTiedCells} evaluated workload--metric
comparisons.  Relative to the best evaluated baseline for each workload and
metric, it reduces peak memory, extra DDR traffic, spill count, and modeled
makespan by \textbf{\OverallPeakVsReference},
\textbf{\OverallDDRVsReference}, \textbf{\OverallSpillVsReference}, and
\textbf{\OverallMakespanVsReference}, respectively.  Plan-preserving CPE
further reduces makespan by \textbf{\CPEGeomeanReduction} over Freeze while
leaving placement and memory traffic unchanged, establishing the static
memory plan as a verifiable scheduling contract between memory planning and
pipeline optimization.
\end{abstract}


\section{Introduction}
\label{sec:introduction}

General-purpose NPUs increasingly execute a mixture of convolution,
matrix multiplication, attention, and generative workloads
~\cite{vaswani2017attention,brown2020gpt3,
dao2022flashattention,dao2023flashattention2}.
To support this range within one core, modern NPUs combine heterogeneous
Cube, Vector, and Scalar engines, multiple memory-transfer engines (MTEs),
and a hierarchy of explicitly managed on-chip buffers
~\cite{tang2023davinci,zhou2025ascend,zhang2026npumeter}.
After lowering, operators are represented as fine-grained command DAGs
distributed across these compute and transfer pipelines.  Decoupled engines
expose compute--transfer overlap, but finite scratchpad capacity forces
physical-address reuse and occasional off-chip movement
~\cite{heo2024neupims,gholami2024memorywall,
lee2023blockgroup,lee2025hopscotch}.
Efficient execution therefore requires the compiler to coordinate command
order, data movement, and physical memory planning rather than optimizing
each in isolation.

\begin{figure*}[!t]
  \centering
  \includegraphics[width=0.96\textwidth]{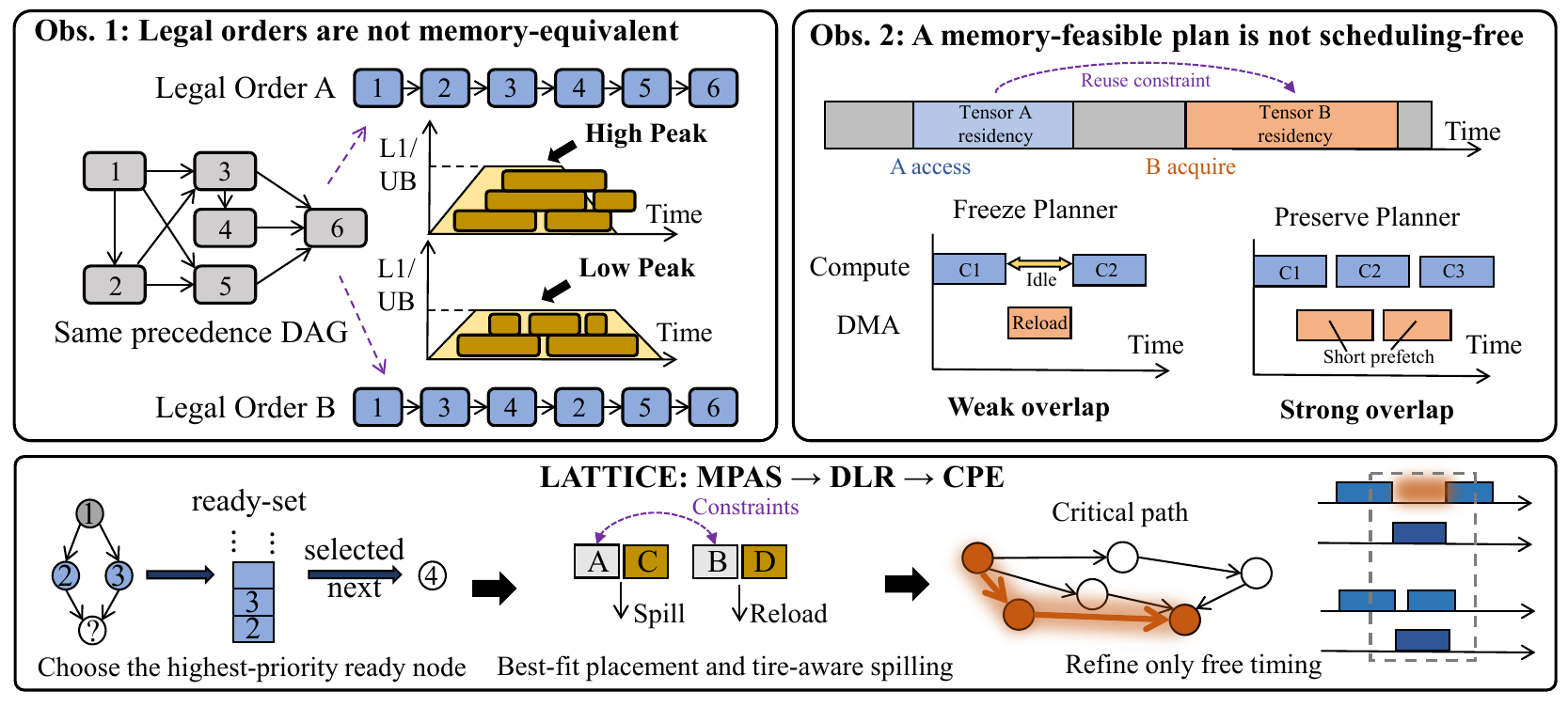}
  \caption{Motivation and \lattice overview.
  Legal order changes lifetime overlap and memory behavior, while
  physical-address reuse introduces constraints absent from the input DAG\@.
  \mpas shapes lifetime geometry, \dlr materializes the memory plan, and
  \cpe refines timing while preserving that plan.}
  \label{fig:lattice-intro}
\end{figure*}

Figure~\ref{fig:lattice-intro} summarizes the two inter-stage effects that
motivate \lattice.

The coupling follows a directed constraint chain.
Command order first determines when each object is allocated, accessed, and
released, thereby shaping the lifetime geometry exposed to memory planning.
The planner then binds these lifetimes to physical intervals and inserts
spill/reload events.  Once materialized, these decisions restrict subsequent
timing: two residencies that reuse the same physical interval cannot own it
concurrently, and an access cannot execute before the event that establishes
its residency.  Thus, order shapes the feasible memory plan, and the realized
memory plan in turn defines the legal space for pipeline refinement.

\textbf{Observation 1: topological legality does not imply memory
equivalence.}
Different legal orders of the same command DAG preserve every original
precedence edge but change allocation-to-release overlap.  They consequently
produce different logical peaks, placement opportunities, spill decisions,
and off-chip traffic.  Under an otherwise identical downstream backend, four
deterministic legal orders exhibit a \textbf{\OrderLogicalRange} per-workload
range in logical L1+UB peak and a \textbf{\OrderDDRRange} range in extra DDR
traffic.  A legal order is therefore not merely an interchangeable
serialization of a fixed memory behavior; it selects the lifetime geometry
from which the physical plan is constructed.  Prior memory-aware scheduling
methods exploit this pre-placement ordering freedom, but a complementary
question arises after planning: once one order has been materialized into
physical addresses and spill/reload epochs, subsequent timing passes can no
longer reason from the original DAG alone.

\textbf{Observation 2: memory feasibility does not imply scheduling
freedom.}
When two physical residencies reuse an overlapping address interval, the
earlier residency's final accesses and closing spill or release must precede
acquisition of the later residency.  These plan-induced constraints may span
Cube, Vector, and MTE pipelines even when the corresponding commands are
independent in the input DAG\@.  Removing only these reuse constraints produces
\textbf{\IgnoreViolationTotal} ownership violations: the resulting schedules
appear faster but are not executable.  Conversely, freezing the complete
planner order is safe but retains resource serialization unrelated to memory
ownership.  The legal timing space therefore lies between these two extremes:
it must preserve the materialized memory plan while releasing the remaining
pipeline-order freedom.

\lattice exposes this missing inter-stage interface by treating the static
memory plan as a first-class scheduling contract.  Following the direction in
which the constraints arise, \mpas first selects a legal order and reshapes
lifetime overlap before address binding; \dlr then materializes the resulting
lifetime geometry into tiered placement, spill/reload events, and plan-induced
reuse constraints; and \cpe refines only the remaining resource order while
preserving the selected memory plan.  This constraint-directed decomposition
allows scheduling, memory planning, and timing refinement to optimize their
respective decision spaces without requiring one monolithic search or
sacrificing execution validity.

This paper makes three contributions:
\begin{itemize}
  \item \textbf{Plan-dependent execution semantics.}
  We identify and quantify both directions of the inter-stage dependency:
  legal command orders induce different lifetime and memory behavior, while
  physical-address reuse introduces execution constraints absent from the
  input precedence DAG\@.

  \item \textbf{Constraint-directed NPU compilation.}
  We present \lattice, a deterministic MPAS--DLR--CPE pipeline in which DLR
  exports an explicit memory-plan contract and CPE recovers compute--transfer
  overlap without invalidating the selected placement or spill/reload
  behavior.

  \item \textbf{End-to-end gains with independent validity evidence.}
  Among four evaluated baselines, \lattice achieves the best or tied-best
  result in all \OverallBestOrTiedCells\ workload--metric comparisons.
  Relative to the best evaluated external method for each workload and
  metric, it reduces peak memory, extra DDR traffic, spill count, and modeled
  makespan by \OverallPeakVsReference, \OverallDDRVsReference,
  \OverallSpillVsReference, and \OverallMakespanVsReference, respectively.
  Controlled Ignore/Freeze/CPE diagnostics and an independent verifier
  establish that these gains come from executable, plan-preserving schedules.
\end{itemize}

\section{Background and Motivation}
\label{sec:background}

\begin{figure*}[!t]
  \centering
  \includegraphics[width=0.96\textwidth]{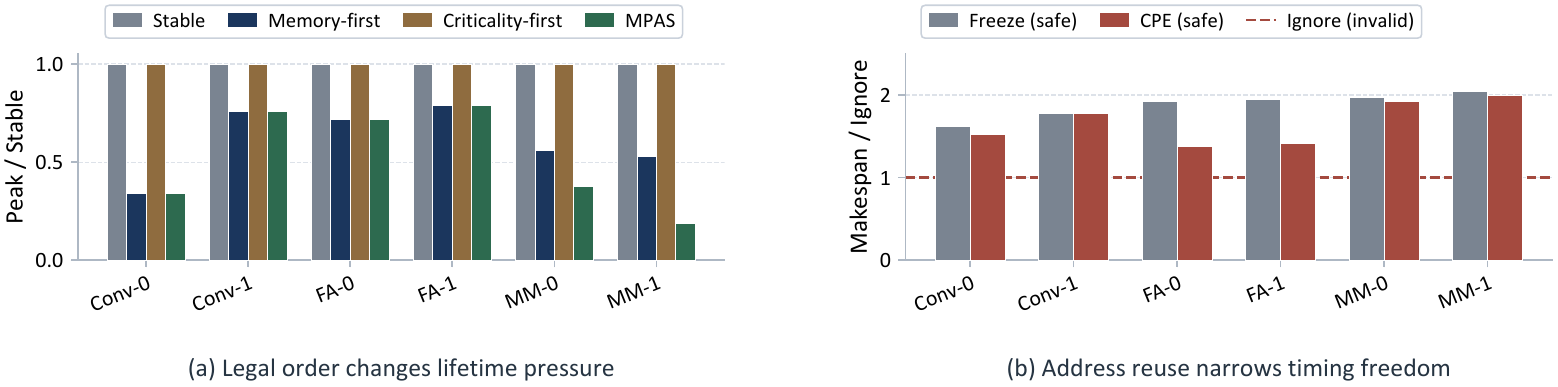}
  \caption{Inter-stage effects.
  (a) Different legal orders of the same command DAG induce different
  logical L1+UB peaks under a common backend.
  (b) Removing plan-induced reuse constraints yields infeasible timing;
  Freeze and CPE preserve the realized memory plan.}
  \label{fig:motivation-profile}
\end{figure*}

\subsection{Intra-Core Execution Model}

\begin{figure*}[!t]
  \begin{minipage}[t]{0.465\textwidth}
    \vspace{0pt}
    \centering
    \includegraphics[pagebox=cropbox,width=\linewidth]{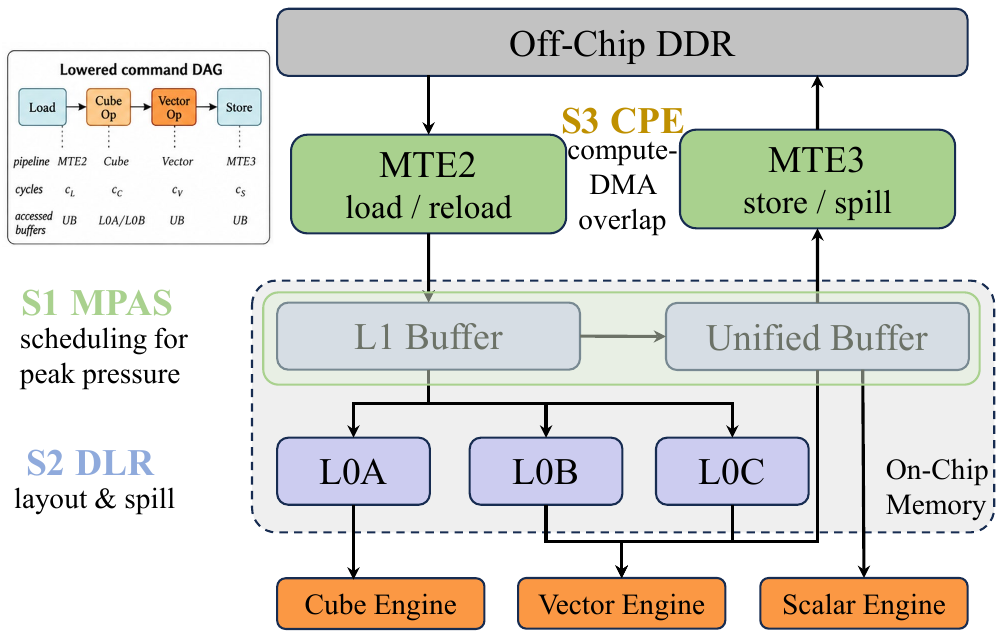}
    \caption{Target architecture and \lattice workflow: \mpas shapes
    lifetimes, \dlr plans memory, and \cpe overlaps compute and DMA.}
    \label{fig:architecture-method-map}
  \end{minipage}
  \hfill
  \begin{minipage}[t]{0.505\textwidth}
    \vspace{0pt}
    \centering
    \makeatletter\def\@captype{table}\makeatother
    \caption{Notation used in formulation and algorithms.}
    \label{tab:notation}
    \small
    \begin{tabular}{@{}p{0.31\linewidth}p{0.62\linewidth}@{}}
      \toprule
      Symbol & Meaning \\
      \midrule
      $G=(V,E_{\mathrm{prec}})$
        & Command DAG with precedence edges \\
      $v,q(v),c_v$
        & Command, assigned pipeline, and execution latency \\
      $b,s_b,\tau(b)$
        & Buffer, size, and home tier \\
      $\sigma,I_b(\sigma)$
        & Legal order and induced lifetime interval \\
      $M_{\max}$
        & Logical L1/UB live peak before placement \\
      $r,x(r)$
        & Physical residency and assigned
          $(\mathrm{tier},\mathrm{offset},\mathrm{size})$ \\
      $\Lambda$
        & Lifecycle and spill/reload event plan \\
      $M_{\mathrm{res}}$
        & Physical resident peak after placement \\
      $P_{\mathrm{mem}}$
        & Fixed memory-plan contract \\
      $E_{\mathrm{fixed}}$
        & Immutable precedence, spill, and reuse constraints \\
      $T_{\mathrm{exec}}$
        & Memory-valid modeled makespan \\
      \bottomrule
    \end{tabular}
  \end{minipage}
\end{figure*}

Figure~\ref{fig:architecture-method-map} summarizes the Da Vinci-derived
target core. General-purpose NPUs integrate heterogeneous compute and
data-movement pipelines within each core: Cube executes matrix-intensive
commands, Vector handles element-wise and reduction operations, Scalar
performs lightweight control, and memory-transfer engines (MTEs) move data
through an explicitly managed memory hierarchy
~\cite{tang2023davinci,zhou2025ascend}. DDR provides off-chip backing
storage; L1 and UB serve as relatively large on-chip staging buffers; and
L0A/L0B/L0C are smaller operand and result scratchpads close to the compute
engines. In our model, MTE2 performs loads and reloads, whereas MTE3 performs
stores and spill-outs.

Unlike a transparent cache hierarchy, these buffers are managed explicitly by
the compiler. Each pipeline is single issue, but commands assigned to different
pipelines may execute concurrently. Performance therefore depends not only on
the amount of computation, but also on when operands become resident, how long
they occupy on-chip storage, and whether data movement can overlap with
computation. Finite capacity further requires different logical objects to
reuse physical address ranges, coupling pipeline timing with memory placement.

The compiler input is a lowered command DAG
$G=(V,E_{\mathrm{prec}})$. Each vertex records its operation, assigned
pipeline, modeled latency, accessed-buffer set, and any ALLOC/FREE action;
buffer metadata provides size and home tier.
$E_{\mathrm{prec}}$ contains the mandatory data and execution precedence
known before memory planning. A logical buffer remains live from allocation
through its final access and release, while spilling and reloading may later
split this logical lifetime into multiple physical residencies.

Physical addresses and residency epochs become explicit downstream of the
input DAG, after the compiler selects a legal command order and
constructs a finite-capacity tiered memory plan. We hold the lowered DAG fixed
and study this low-level boundary, where command ordering first shapes memory
behavior and the resulting physical plan subsequently restricts timing
optimization.

\subsection{Observation 1: Topological Legality Does Not Imply Memory
Equivalence}

A legal topological order preserves every edge in
$E_{\mathrm{prec}}$, yet changes when buffers are allocated, accessed, and
released. It therefore changes lifetime overlap, peak logical pressure, the
address holes exposed to placement, and the victim sequence selected during
spilling. The same input DAG can consequently induce different physical
memory plans before any hardware parameter or data dependency changes.

Figure~\ref{fig:motivation-profile}(a) compares Stable, Memory-first,
Criticality-first, and MPAS under the same downstream DLR/CPE backend.
Changing the legal order produces a
\textbf{\OrderLogicalRange} per-workload range in logical L1+UB peak and a
\textbf{\OrderDDRRange} range in extra DDR traffic. Thus, a legal order is
a first-class memory-planning decision that determines the lifetime geometry
presented to the downstream planner.

Prior memory-aware schedulers exploit this upstream ordering freedom
~\cite{ahn2020ordering,kumar2019efficient,lee2023blockgroup,
wen2020timespace,yu2026opass}. LATTICE builds on this insight but focuses on
the complementary downstream question: after one order has been materialized
as a concrete physical memory plan, which timing transformations remain legal?

\subsection{Observation 2: Memory Feasibility Does Not Imply Scheduling
Freedom}

A physical memory plan couples tier assignments and offsets with ownership
epochs.
If two residencies reuse an overlapping address interval, every access and
closing spill or release of the earlier residency must complete before the
later residency acquires that interval. Such a handoff can connect Cube,
Vector, and MTE commands that are independent in the input DAG\@. Physical
address reuse therefore introduces plan-induced ordering constraints that
cannot be inferred from $E_{\mathrm{prec}}$ alone.

The selected plans contain \textbf{\ReuseEdgeTotal} strict access-level
reuse constraints. Figure~\ref{fig:motivation-profile}(b) removes only these
constraints in \emph{Ignore}, while retaining input precedence, lifecycle,
transfer, and pipeline-resource constraints. The resulting timings appear
shorter, but produce \textbf{\IgnoreViolationTotal} ownership conflicts and
are rejected by the independent verifier. This diagnostic confirms that
topological legality alone does not preserve an already materialized memory
plan.

\emph{Freeze} provides a verified reference by preserving the complete
Stage-2 resource order, including serialization beyond residency and address
ownership requirements. CPE preserves the plan-induced constraints while
reducing geometric-mean makespan by \textbf{\CPEGeomeanReduction} relative
to Freeze.
Post-placement timing freedom is therefore neither the unconstrained space of
the input DAG nor the fully frozen planner order; it is the space left legal
by the realized memory plan.

\subsection{Our Approach}

The two observations expose a directed compiler interface:
\emph{command order shapes lifetime geometry, memory planning materializes
that geometry into physical execution constraints, and timing refinement must
inherit those constraints}.  \lattice realizes this interface through three
deterministic stages.  \mpas selects a memory-aware legal order before address
binding; \dlr converts the resulting lifetime geometry into tiered placement,
lifecycle and spill/reload events, and plan-induced reuse constraints; and
\cpe refines only the resource-order freedom left by the fixed memory plan,
accepting candidates that improve timing and pass independent verification.

This constraint-directed organization avoids a monolithic search over
scheduling, placement, and timing while preserving end-to-end validity.
The materialized memory plan becomes a first-class compiler interface,
allowing each stage to optimize its own decision space under a shared legality
boundary.  Figure~\ref{fig:architecture-method-map} shows where \mpas, \dlr,
and \cpe act across the command, memory, and compute--transfer layers.

\section{Problem Formulation}
\label{sec:formulation}

Table~\ref{tab:notation} summarizes the notation used throughout the
formulation and algorithms.

\subsection{Order-Dependent Lifetime Semantics}

We model the lowered execution as a command DAG
$G=(V,E_{\mathrm{prec}})$.
Each command $v$ is associated with a pipeline $q(v)$, execution latency
$c_v$, and accessed buffers. Before physical placement, the compiler can
select any legal topological order:
\begin{equation}
\mathcal{T}(G)=
\{\sigma\mid \sigma(u)<\sigma(v),
\forall (u,v)\in E_{\mathrm{prec}}\}.
\end{equation}

For a buffer $b$ with size $s_b$ and home tier $\tau(b)$, let $a_b$ and
$f_b$ denote its allocation and release commands. A legal order determines
its lifetime interval:
\begin{equation}
I_b(\sigma)
=
[\sigma(a_b),\sigma(f_b)]\cap\mathbb{Z}.
\end{equation}

Therefore, command ordering determines the lifetime geometry exposed to
memory planning. The logical footprint of tier $m$ at position $k$ is
\begin{equation}
R_m(k;\sigma)
=
\sum_{b:\tau(b)=m}
s_b\mathbf{1}[k\in I_b(\sigma)],
\end{equation}
and the pre-placement upper-tier peak is
\begin{equation}
M_{\max}(\sigma)
=
\max_k
\sum_{m\in\{\mathrm{L1},\mathrm{UB}\}}
R_m(k;\sigma).
\end{equation}

\subsection{Memory Plan Contract}

A memory planner converts the selected lifetime structure into a physical
plan. Due to spilling and reloading, one logical buffer may generate
multiple physical residencies. Each residency $r$ is assigned:
\begin{equation}
x(r)=(m_r,o_r,s_r).
\end{equation}

A valid placement requires that simultaneously active residencies in the
same tier occupy disjoint address ranges:
\begin{align}
&\mathrm{TimeOverlap}(r_i,r_j)\land m_{r_i}=m_{r_j}
\nonumber\\
&\Rightarrow
[o_i,o_i+s_i)\cap[o_j,o_j+s_j)=\varnothing .
\end{align}

Let $J_r$ be the acquire-to-release interval of residency $r$ in the
replayed plan. The resulting physical resident peak is:
\begin{equation}
M_{\mathrm{res}}
=
\max_k
\sum_{r:m_r\in\mathcal M_{\mathrm{on}}}
s_r\mathbf1[k\in J_r].
\end{equation}

After placement, the selected layout $x^\star$, event stream
$\Lambda^\star$, and physical-address reuse relations
$E_{\mathrm{reuse}}$ are fixed. We define the memory-plan contract as:
\begin{equation}
P_{\mathrm{mem}}
=
(x^\star,\Lambda^\star,E_{\mathrm{reuse}}).
\label{eq:memory-contract}
\end{equation}

This contract specifies the memory behavior that subsequent timing
refinement must preserve.

\subsection{Plan-Constrained Timing}

Spill, reload, and lifecycle events are promoted to timing vertices:
\begin{equation}
V'=V\cup V_{\Lambda}.
\end{equation}

The immutable constraints induced by the original DAG and memory plan are:
\begin{equation}
E_{\mathrm{fixed}}
=
E_{\mathrm{prec}}
\cup
E_{\mathrm{spill}}
\cup
E_{\mathrm{reuse}} .
\label{eq:fixed-edges}
\end{equation}

A timing refinement is expressed through resource-order constraints:
\begin{equation}
G_{\mathrm{aug}}(\rho)
=
(V',E_{\mathrm{fixed}}\cup E_{\mathrm{pipe}}(\rho)).
\end{equation}

Let $\theta_v$ and $\phi_v=\theta_v+c_v$ denote the start and finish time
of command $v$. A schedule is valid when it satisfies the fixed graph,
memory feasibility, and pipeline resource constraints. The corresponding
memory-valid modeled execution time is:
\begin{equation}
T_{\mathrm{exec}}
=
\max_{v\in V'}\phi_v .
\end{equation}

Based on this formulation, LATTICE implements three deterministic stages:
MPAS selects an order that shapes lifetime geometry, DLR materializes the
memory-plan contract, and CPE optimizes timing over the feasible region
defined by that contract.

\begin{figure*}[t]
  \centering
  \includegraphics[width=0.96\textwidth]{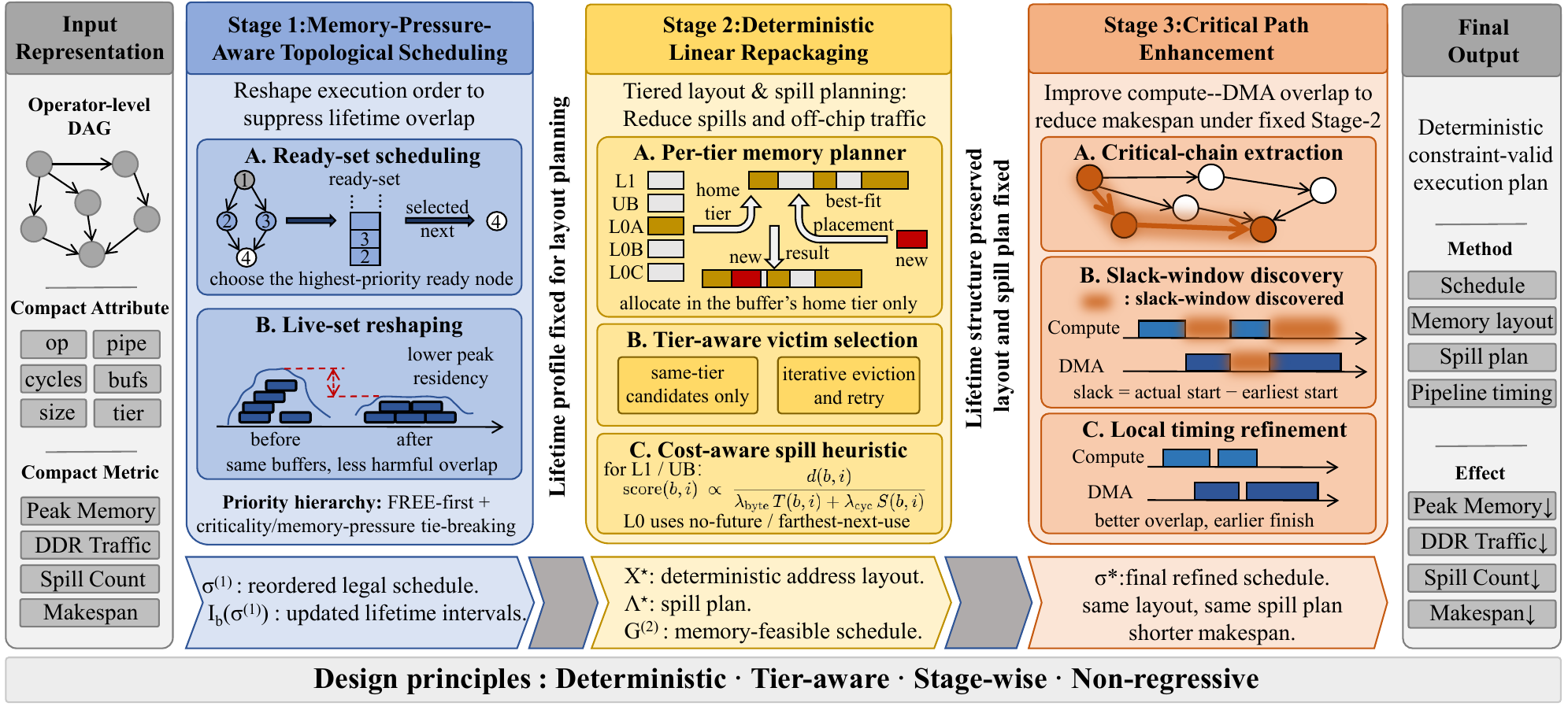}
  \caption{Constraint-directed \lattice flow.
  \mpas selects a legal order and reshapes lifetime geometry before address
  binding. \dlr materializes tiered placement, spill/reload events, and
  plan-induced reuse constraints. \cpe refines resource order while
  preserving the resulting memory plan.}
  \label{fig:overview}
\end{figure*}

\section{LATTICE Design}
\label{sec:design}

\subsection{Overview}
\label{subsec:design-overview}

Figure~\ref{fig:overview} shows the three-stage \lattice pipeline.
Stage~1, \emph{Memory-Pressure-Aware Topological Scheduling} (\mpas),
selects a legal command order $\sigma^{(1)}$ and reshapes the lifetime
geometry exposed to memory planning. Stage~2, \emph{Deterministic Linear
Repackaging} (\dlr), binds this geometry to tiered addresses and explicit
spill/reload events, producing the fixed memory-plan contract
\begin{equation}
P_{\mathrm{mem}}
=
(x^\star,\Lambda^\star,E_{\mathrm{reuse}}).
\label{eq:design-memory-contract}
\end{equation}
Here, $x^\star$ is the physical layout, $\Lambda^\star$ is the lifecycle and
spill/reload event plan, and $E_{\mathrm{reuse}}$ captures the ordering
required by physical-address reuse. Stage~3, \emph{Critical Path
Enhancement} (\cpe), refines the resource-order edges
$E_{\mathrm{pipe}}(\rho)$ subject to the fixed memory plan.

The stage boundaries therefore encode explicit compiler semantics:
\mpas passes lifetime geometry to \dlr; \dlr materializes the memory
semantics that later scheduling must preserve; and \cpe extracts the timing
freedom permitted by that plan. Every accepted output passes the independent
memory and timing verifier described in Section~\ref{subsec:validity}.

\subsection{Memory-Pressure-Aware Topological Scheduling}
\label{subsec:mpas}

\mpas exploits the fact that different legal orders induce different
allocation-to-release overlap. It performs a deterministic Kahn traversal
and selects the next command only from the current ready set.

For a lifecycle command $v$, let $b(v)$ denote its associated buffer. We
define
\begin{align}
L(v)&=
\begin{cases}
0, & v\text{ is FREE},\\
1, & v\text{ is an ordinary command},\\
2, & v\text{ is ALLOC},
\end{cases}
\label{eq:mpas-lifecycle}
\\
\Delta(v)&=
\begin{cases}
-s_{b(v)}, & v\text{ is FREE},\\
+s_{b(v)}, & v\text{ is ALLOC},\\
0, & \text{otherwise}.
\end{cases}
\label{eq:mpas-pressure}
\end{align}
The lifecycle class advances releases and delays allocations, while
$\Delta(v)$ favors larger releases and smaller allocations within the same
class. To avoid unnecessarily delaying long dependency chains, \mpas also
uses the downstream bottom level
\begin{equation}
C(v)
=
c_v+\max_{w\in\mathrm{succ}(v)}C(w),
\label{eq:mpas-criticality}
\end{equation}
where the maximum is zero for a sink. The next ready command minimizes
\begin{equation}
K_{\mathrm{MPAS}}(v)=
\left(
L(v),\Delta(v),-C(v),
\mathrm{inputpos}(v),\mathrm{id}(v)
\right)
\label{eq:mpas-key}
\end{equation}
lexicographically. The final two fields provide stable tie breaking.

Because commands are selected only after all predecessors have been issued,
the resulting $\sigma^{(1)}$ is legal by construction. \mpas determines only
command order and the induced lifetime intervals; physical placement and
spill decisions remain deferred to \dlr.

\begin{algorithm}[t]
\caption{Memory-Pressure-Aware Topological Scheduling}
\label{alg:mpas}
\begin{algorithmic}[1]
\REQUIRE Command DAG $G=(V,E_{\mathrm{prec}})$
\ENSURE Legal deterministic order $\sigma^{(1)}$
\STATE Compute $C(v)$ for all $v$ in reverse topological order
\STATE Initialize indegrees and ready heap
       $Q\leftarrow\{v\mid\mathrm{indeg}(v)=0\}$
       using Eq.~\eqref{eq:mpas-key}
\STATE $\sigma^{(1)}\leftarrow[~]$
\WHILE{$Q\neq\emptyset$}
  \STATE $v\leftarrow\mathrm{popmin}(Q)$; append $v$ to $\sigma^{(1)}$
  \FORALL{$w\in\mathrm{succ}(v)$}
    \STATE $\mathrm{indeg}(w)\leftarrow\mathrm{indeg}(w)-1$
    \IF{$\mathrm{indeg}(w)=0$}
      \STATE Insert $w$ into $Q$
    \ENDIF
  \ENDFOR
\ENDWHILE
\STATE Reject unless $\sigma^{(1)}$ contains every vertex exactly once
\RETURN $\sigma^{(1)}$
\end{algorithmic}
\end{algorithm}

\subsection{Deterministic Linear Repackaging}
\label{subsec:dlr}

Given $\sigma^{(1)}$, \dlr scans the command stream and maintains an
offset-ordered free list and resident map for every memory tier. For a
requested residency of size $s_b$ in its home tier $\tau(b)$, the feasible
holes are
\begin{equation}
\mathcal C_b
=
\left\{
[o,o+\ell)\in\mathcal F_{\tau(b)}
\mid \ell\geq s_b
\right\}.
\end{equation}
When $\mathcal C_b$ is nonempty, deterministic best fit selects
\begin{equation}
[o^\star,o^\star+\ell^\star)
=
\arg\min_{[o,o+\ell)\in\mathcal C_b}
\left(\ell-s_b,\;o\right),
\label{eq:dlr-best-fit}
\end{equation}
where the comparison is lexicographic. Operands of the current command are
pinned while its memory state is prepared, preventing their eviction.

When no feasible interval exists, \dlr selects a victim from the unpinned
residencies in the requested tier. Let $\nu_b(k)$ be the next access to
buffer $b$ after scan position $k$, let $\ell_b$ be its final access
position, and define the implemented writeback cost
\begin{equation}
W_b=(\alpha+\beta s_b)+s_b .
\end{equation}
For L1 and UB, victims are ranked by
\begin{equation}
K_{\mathrm{victim}}(b,k)=
\left(
\mathbf 1_{\mathrm{noWB}}(b),
\frac{\nu_b(k)-k}{\max(1,W_b)},
-\ell_b,
-\mathrm{id}_b
\right),
\label{eq:dlr-victim-key}
\end{equation}
and the lexicographically largest key is selected.
$\mathbf 1_{\mathrm{noWB}}(b)$ marks a dead or safely rematerializable
residency, and $\nu_b(k)=+\infty$ when no future access exists. The policy
therefore avoids unnecessary writeback first and otherwise balances reuse
distance with spill cost. LRU is retained as the controlled alternative in
the Stage-2 ablation. For L0A/L0B/L0C, \dlr applies eager flushing to
unpinned near-compute residencies, preferring no-future-use and then
farthest-next-use victims.

A victim is dropped when a future materialization can reconstruct it;
otherwise, \dlr emits an explicit SPILL\_OUT\@. A later access emits
SPILL\_IN only when a valid backing copy exists. Each allocation,
materialization, spill, reload, drop, and release is anchored to a command
phase, producing the event plan $\Lambda^\star$ and the associated
residency epochs.

\paragraph{Memory-plan contract.}
The Stage-2 output is more than an offset assignment. Acquisition, access,
spill, reload, and release relations form $E_{\mathrm{spill}}$, which bounds
each access by a valid residency and backing epoch. If two same-tier
residencies reuse overlapping physical intervals, \dlr orients their
ownership handoff according to the realized plan and expands it into
explicit access-level ownership barriers $E_{\mathrm{reuse}}$. It then constructs
\begin{equation}
G_{\mathrm{fixed}}
=
\left(
V',
E_{\mathrm{prec}}\cup E_{\mathrm{spill}}\cup E_{\mathrm{reuse}}
\right).
\label{eq:design-fixed-graph}
\end{equation}
Thus, \dlr converts the lifetime geometry selected by \mpas into the physical
layout, event plan, and fixed dependencies consumed by \cpe.

\begin{algorithm}[t]
\caption{Deterministic Linear Repackaging}
\label{alg:dlr}
\begin{algorithmic}[1]
\REQUIRE $G$, legal order $\sigma^{(1)}$, tier capacities $\mathbf C$
\ENSURE Fixed memory-plan contract $P_{\mathrm{mem}}$
\STATE Initialize one free list and resident map per tier
\FOR{command $v$ in $\sigma^{(1)}$}
  \STATE Apply the ALLOC/FREE action associated with $v$
  \STATE Pin all operands required by $v$
  \FORALL{required object $b$ that is not resident}
    \WHILE{no best-fit interval exists for $b$}
      \STATE Select a deterministic same-tier unpinned victim
      \STATE Emit DROP or SPILL\_OUT and release its interval
    \ENDWHILE
    \STATE Assign the best-fit interval
    \STATE Emit ALLOCATE, MATERIALIZE, or SPILL\_IN as required
  \ENDFOR
  \STATE Record accesses and update backing-state validity
\ENDFOR
\STATE Derive address-reuse ownership order
\STATE Construct $E_{\mathrm{spill}}$, $E_{\mathrm{reuse}}$,
       and $G_{\mathrm{fixed}}$
\RETURN $(x^\star,\Lambda^\star,E_{\mathrm{reuse}})$
\end{algorithmic}
\end{algorithm}

\subsection{Plan-Preserving Critical Path Enhancement}
\label{subsec:cpe}

\cpe starts from the complete safe Stage-2 resource order $\rho_0$, denoted
\emph{Freeze}. Freeze preserves the memory plan but may retain
single-pipeline serialization that is not required by
$E_{\mathrm{fixed}}$.

Let $\underline{\theta}_v$ be the earliest start time of command $v$ under
$G_{\mathrm{fixed}}$, and let $\theta_v^0$ be its start time under Freeze.
The resource-order slack is
\begin{equation}
\mathrm{slack}(v)
=
\theta_v^0-\underline{\theta}_v .
\label{eq:cpe-slack}
\end{equation}
Positive slack identifies delay introduced by pipeline order rather than by
the fixed memory constraints.

\cpe first constructs one deterministic list-scheduling seed over
$G_{\mathrm{fixed}}$, using predicted finish time, downstream criticality,
slack, resource idle gap, and stable input order. The seed exclusively refines
$E_{\mathrm{pipe}}(\rho)$ and is retained when it preserves
$P_{\mathrm{mem}}$, passes verification, and improves Freeze.

Starting from the better valid order, \cpe extracts one critical chain and
enumerates adjacent same-pipeline swaps within a bounded neighborhood of that
chain. A candidate $\rho'$ is accepted only if:

\begin{enumerate}
  \item $G_{\mathrm{aug}}(\rho')$ is acyclic and preserves
        $E_{\mathrm{fixed}}$;
  \item every real pipeline remains single issue;
  \item $x^\star$, $\Lambda^\star$, reuse orientation, physical resident
        peak, spill count, and DDR traffic remain unchanged;
  \item the independent verifier accepts the execution; and
  \item $T_{\mathrm{exec}}(\rho')<T_{\mathrm{exec}}(\rho)$.
\end{enumerate}

The search stops at the fixed pass bound or when no valid move improves
makespan. Freeze is returned as the non-regressing fallback. Therefore,
\cpe recovers beneficial overlap within its bounded neighborhood while
preserving every memory-induced dependency of the materialized plan.

\begin{algorithm}[t]
\caption{Plan-Preserving Critical Path Enhancement}
\label{alg:cpe}
\begin{algorithmic}[1]
\REQUIRE $G_{\mathrm{fixed}}$, $P_{\mathrm{mem}}$,
         frozen resource order $\rho_0$
\ENSURE Verified non-regressing order $\widehat{\rho}$
\STATE $\rho\leftarrow\rho_0$; evaluate verified makespan $T$
\STATE Compute fixed-graph criticality and slack
\STATE Construct deterministic slack-guided seed $\rho_s$
\IF{$\rho_s$ preserves $P_{\mathrm{mem}}$, passes verification,
    and $T(\rho_s)<T$}
  \STATE $\rho\leftarrow\rho_s$; $T\leftarrow T(\rho_s)$
\ENDIF
\FOR{$p=1$ to the bounded local-pass count}
  \STATE Extract a critical chain from $G_{\mathrm{aug}}(\rho)$
  \STATE Enumerate bounded adjacent swaps near the chain
  \STATE Discard candidates violating fixed edges or the memory plan
  \STATE Select the valid candidate with the smallest $T'$
  \IF{no candidate satisfies $T'<T$}
    \STATE \textbf{break}
  \ENDIF
  \STATE $\rho\leftarrow\rho'$; $T\leftarrow T'$
\ENDFOR
\RETURN $\widehat{\rho}\leftarrow\rho$
\end{algorithmic}
\end{algorithm}

\subsection{Validity, Determinism, and Complexity}
\label{subsec:validity}

A plan-level replay checks command coverage, event anchors, tier capacity,
residency transitions, valid backing copies, operand availability, and
physical address non-overlap. A timing-level replay verifies all edges in
$E_{\mathrm{fixed}}\cup E_{\mathrm{pipe}}(\rho)$, ownership handoffs,
single-issue execution, and graph acyclicity. Both replays recompute physical
resident peak, DDR traffic, and spill count; an accepted CPE result must match
the Stage-2 plan on all three metrics.

Determinism follows from fixed metadata and capacities, lexicographic keys,
stable traversal order, deterministic placement and victim selection,
bounded candidate enumeration, and strict-improvement acceptance.

\mpas runs in
$O((|V|+|E_{\mathrm{prec}}|)\log |V|)$ with a ready heap.
\dlr performs one command/event scan plus per-tier free-list and victim
operations. Fixed-edge construction depends on the number of reuse-related
residency and access pairs. With at most $K$ evaluated timing candidates,
\cpe costs
\begin{equation}
O\!\left(K\left(|V'|+|E_{\mathrm{aug}}|\right)\right),
\end{equation}
where $K$ is bounded by the configured seed and local-search limits.
All costs are incurred offline; disabling \cpe retains the verified
MPAS+\dlr Freeze plan.

\section{Experimental Setup}
\label{sec:setup}

We evaluate end-to-end compiler-policy baselines and controlled \lattice
variants over the same low-level command traces and target execution model.  The
end-to-end comparison preserves each method's ordering and memory-policy
decisions, whereas the controlled studies share the \lattice backend to
isolate individual stages and constraints.

\subsection{Workloads and Baselines}
\label{subsec:workloads}

\paragraph{Workloads.}
We use six artifact-provided command DAGs from Da Vinci-derived NPU execution
flows: Case0 and Case1 traces for convolution,
FlashAttention, and matrix multiplication.  The common loader reconstructs
82,470 commands, 165,562 input-precedence edges, and 25,920 explicit
allocations totaling 2,377,703 trace storage units.  Each command records its
operation, execution pipeline, modeled latency, accessed buffers, and
ALLOC/FREE action; buffer metadata provides size and home tier.

\begin{figure*}[t]
  \centering
  \includegraphics[width=0.98\textwidth]
  {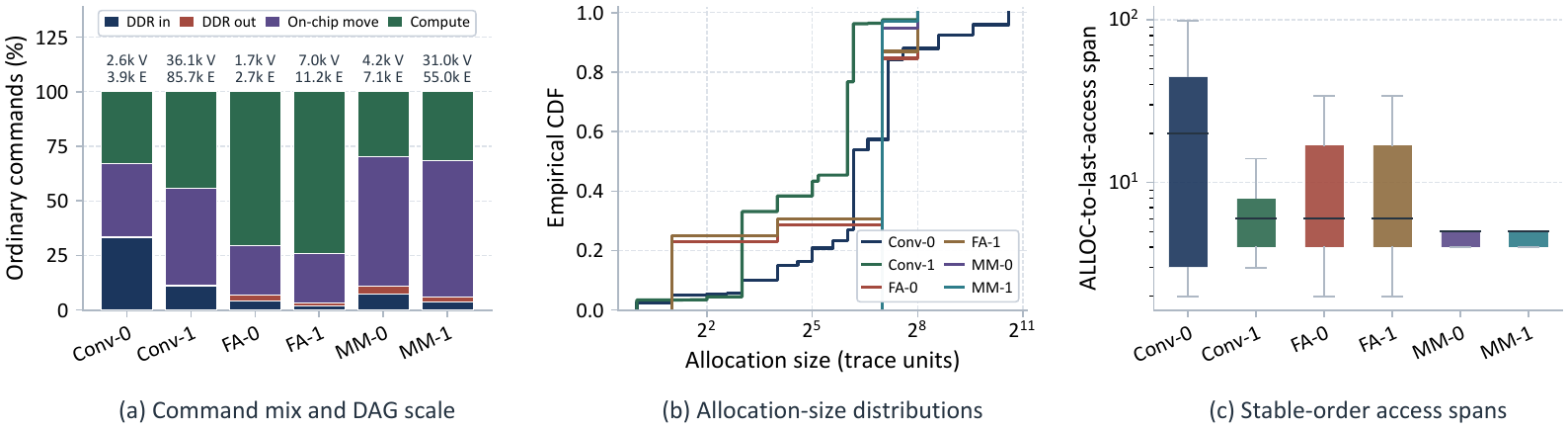}
  \caption{Workload characterization.
  (a) Command composition and DAG scale, excluding ALLOC/FREE commands.
  (b) Allocation-size distributions in trace storage units.
  (c) ALLOC-to-last-access spans in the stable input order.
  $V$ and $E$ denote command vertices and input-precedence edges.}
  \label{fig:workload-characterization}
\end{figure*}

Figure~\ref{fig:workload-characterization} shows substantial variation in
graph scale, command composition, allocation granularity, and access span.
The traces range from 1.7K commands and 2.7K edges to 36.1K commands and
85.7K edges.  Convolution mixes DDR movement, on-chip movement, and
computation; FlashAttention is compute-intensive; and matrix multiplication
contains a larger fraction of on-chip movement.  These differences expose
distinct lifetime-overlap, fragmentation, and pipeline-overlap conditions.

The artifact exposes the low-level command, buffer, tier, and timing fields
required for controlled compiler analysis.  Because the source network,
tensor shapes, chip SKU, and compiler version are not encoded, we report the
artifact's native storage units and modeled cycles, keeping every metric tied
to directly observable trace semantics.

\paragraph{End-to-end baselines.}
We compare \lattice with command-level policy instantiations of
HMCOS~\cite{wang2022hmcos}, COSMA~\cite{li2023cosma},
MAGIS~\cite{chen2024magis}, and EDA~\cite{pao2025eda}.
All methods process the same workloads and are evaluated with the same target
tiers, transfer model, timing semantics, and metric definitions, while
retaining their own ordering and memory-policy decisions.

HMCOS follows the scheduling structure exposed by its public implementation;
COSMA instantiates its joint scheduling, allocation, and replacement policy
through the deterministic common backend; MAGIS applies its graph-reordering
policy to the available command DAG; and EDA follows its published traversal
strategy.  These command-level instantiations preserve the defining policy
decisions represented in the shared low-level IR, enabling controlled
end-to-end replay under identical timing and memory semantics.  The comparison
therefore isolates policy behavior from toolchain-specific engineering.
\textbf{Base} is the stable-order reference policy with no \lattice stages and
provides the normalization reference.

\paragraph{Controlled variants.}
Full \textbf{LATTICE} combines MPAS, cost-aware DLR, and plan-preserving CPE\@.
Stage attribution starts from \textbf{Original+LRU+Freeze} and progressively
enables MPAS, the DLR victim policy, and CPE\@.  \textbf{LATTICE w/o CPE}
retains the complete safe Stage-2 resource order, while
\textbf{LATTICE-LRU w/o CPE} additionally replaces DLR's victim policy with
LRU\@.

For order sensitivity, Stable, Memory-first, Criticality-first, and MPAS use
the same DLR/CPE backend.  For constraint diagnostics, \emph{Ignore} removes
$E_{\mathrm{reuse}}$ from a fixed MPAS+DLR plan, \emph{Freeze} retains its
complete Stage-2 resource order, and CPE changes only resource-order edges
while preserving $E_{\mathrm{fixed}}$.  The phase-direction counterfactual
reuses an MPAS+DLR event/address plan under the Stable command order without
replanning.

\subsection{Replay Model and Metrics}
\label{subsec:replay}

Our deterministic compiler-replay prototype comprises a common trace loader,
tiered-memory planning, augmented-graph timing, and the independent verifier
described in Section~\ref{subsec:validity}.  ALLOCATE, FREE, DROP, and other
zero-latency lifecycle events consume no execution pipeline.  SPILL\_OUT and
SPILL\_IN are explicit timed vertices assigned to MTE3 and MTE2,
respectively.  Each real pipeline is single issue, while commands on distinct
pipelines may overlap.  Earliest feasible timing over the augmented graph
determines modeled makespan and compute--MTE overlap.

\begin{table}[t]
\centering
\caption{Default memory and transfer model in trace units.}
\label{tab:configuration}
\begin{tabular}{ll}
\toprule
Item & Value \\
\midrule
On-chip tiers & L1, UB, L0A, L0B, L0C \\
Capacities & 4096, 1024, 256, 256, 512 \\
Spill-out/reload pipeline & MTE3/MTE2 \\
Added-transfer cost & $150+2q_e$ modeled cycles \\
Per-pipeline issue & Single issue \\
Cross-pipeline execution & Overlap permitted \\
CPE local-search bound & 2 passes, at most 4 candidate moves \\
CPE initial candidate & One deterministic slack-guided seed \\
\bottomrule
\end{tabular}
\end{table}

\begin{figure*}[t]
  \centering
  \includegraphics[width=0.98\textwidth]
  {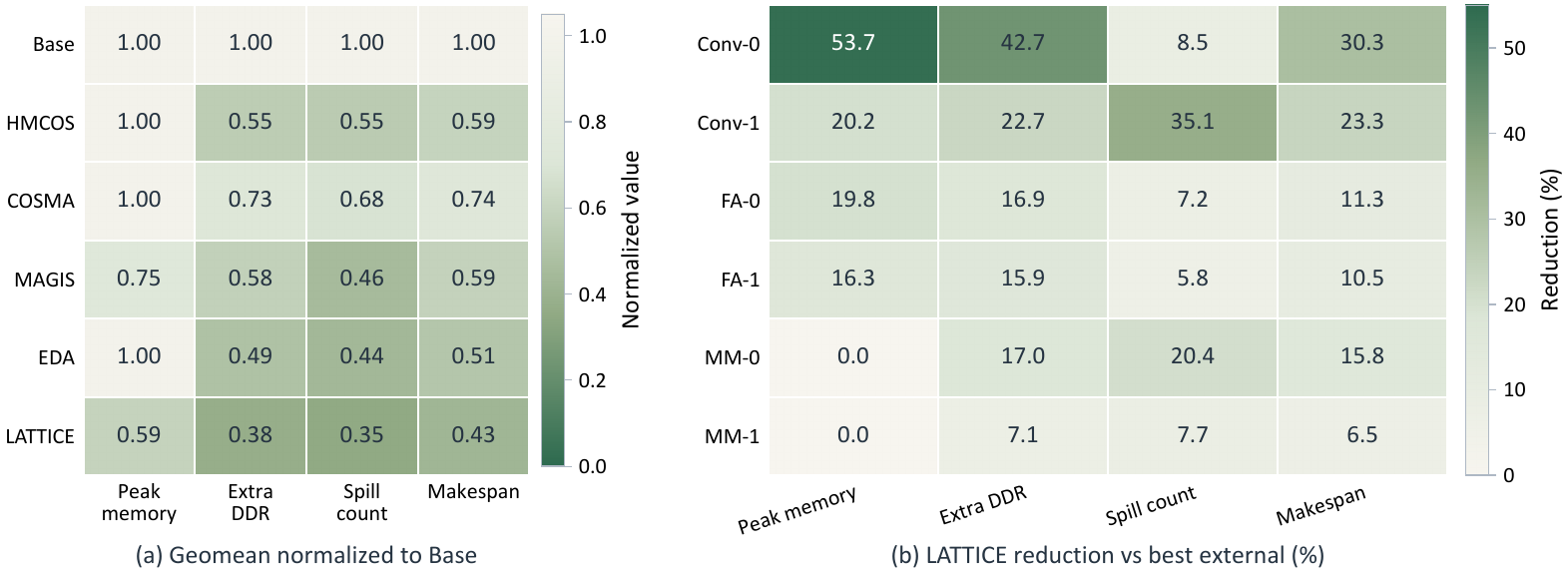}
    \vspace{-10pt}
  \caption{End-to-end comparison under unified replay.
  (a) Geometric means normalized to Base; lower is better.
  (b) \lattice reduction relative to the best evaluated external method
  for each workload and metric; higher is better.}
    \vspace{-10pt}
  \label{fig:overall-performance}
\end{figure*}

All controlled \lattice variants use the configuration in
Table~\ref{tab:configuration}, identical event semantics, and the same
planner, timing backend, and verifier.  An accepted CPE output must preserve
the Stage-2 layout, event plan, physical resident peak, spill count, and DDR
traffic while satisfying all fixed and resource-order constraints.

The end-to-end comparison reports peak memory, planner-introduced two-way DDR
transfer units, spill count (SPILL\_OUT events), and modeled makespan.  Controlled diagnostics
additionally distinguish the pre-placement logical L1+UB peak from the
post-placement physical resident peak across all five on-chip tiers.

Normalized aggregate results use geometric means of per-workload ratios.
Improvements over external methods are arithmetic means of per-workload
percentage reductions relative to the best evaluated external method for the
corresponding workload and metric.  CPE improvement uses the geometric mean
of per-workload CPE/Freeze makespan ratios.  Each percentage identifies its
reference in the accompanying text or caption.

The evaluation focuses on static, single-core command streams to isolate
compiler decisions from device-specific runtime variation.  The replay model
abstracts dynamic arrivals, continuous batching, sub-buffer access phases, and
multi-core compilation; whole-residency ownership supplies a strict,
architecture-neutral safety criterion at this compiler boundary.

\section{Experimental Results}
\label{sec:evaluation}

\subsection{Overall Comparison}
\label{subsec:eval_overall}

Figure~\ref{fig:overall-performance}(a) compares the complete compiler
policies under the unified replay protocol.  Relative to Base, \lattice
reduces peak memory, extra DDR traffic, spill count, and modeled makespan by
\textbf{\OverallPeakVsBase}, \textbf{\OverallDDRVsBase},
\textbf{\OverallSpillVsBase}, and
\textbf{\OverallMakespanVsBase}, respectively.  The coordinated gains across
all four metrics show that the three stages improve memory efficiency and
execution time simultaneously.

Figure~\ref{fig:overall-performance}(b) compares \lattice with the best
evaluated external method separately for each workload and metric.  \lattice
is best or tied-best in all \textbf{\OverallBestOrTiedCells}
workload--metric comparisons.  Averaged across the six workloads, it reduces
peak memory by \textbf{\OverallPeakVsReference}, extra DDR traffic by
\textbf{\OverallDDRVsReference}, spill count by
\textbf{\OverallSpillVsReference}, and makespan by
\textbf{\OverallMakespanVsReference}.  Two matrix-multiplication peak-memory
cells match the strongest external result; all other cells are strict
improvements.  The largest joint gain occurs on Conv-0, where \lattice
reduces peak memory by 53.7\%, extra DDR traffic by 42.7\%, spill count by
8.5\%, and makespan by 30.3\% relative to the corresponding strongest
external results.

\subsection{Stage and Order Analysis}
\label{subsec:eval_stage_order}

\begin{figure*}[t]
  \centering
  \includegraphics[width=0.98\textwidth]
  {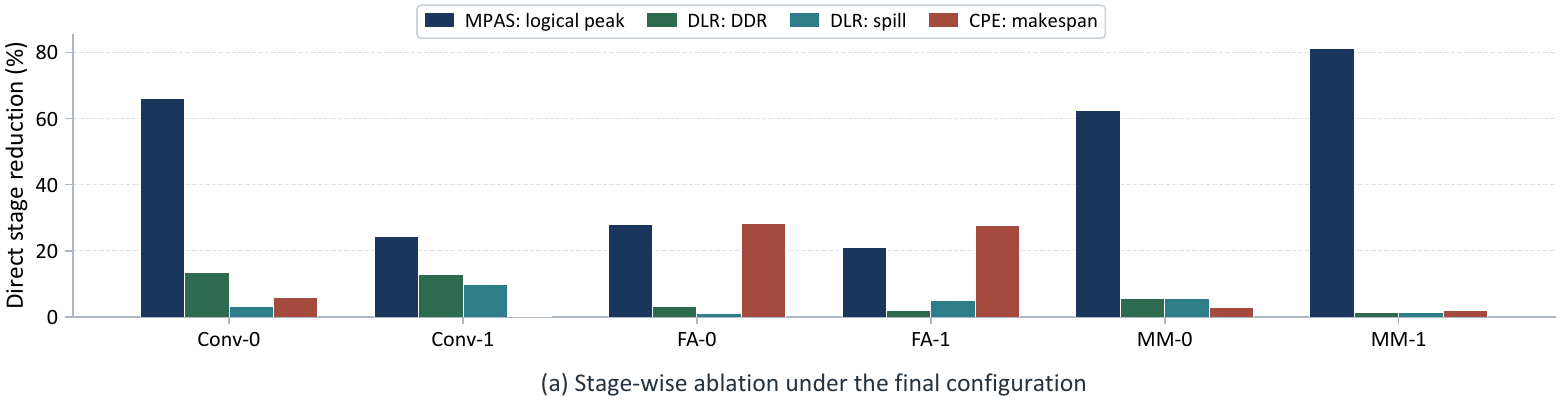}
  \vspace{0.5ex}
  \includegraphics[width=0.98\textwidth]
  {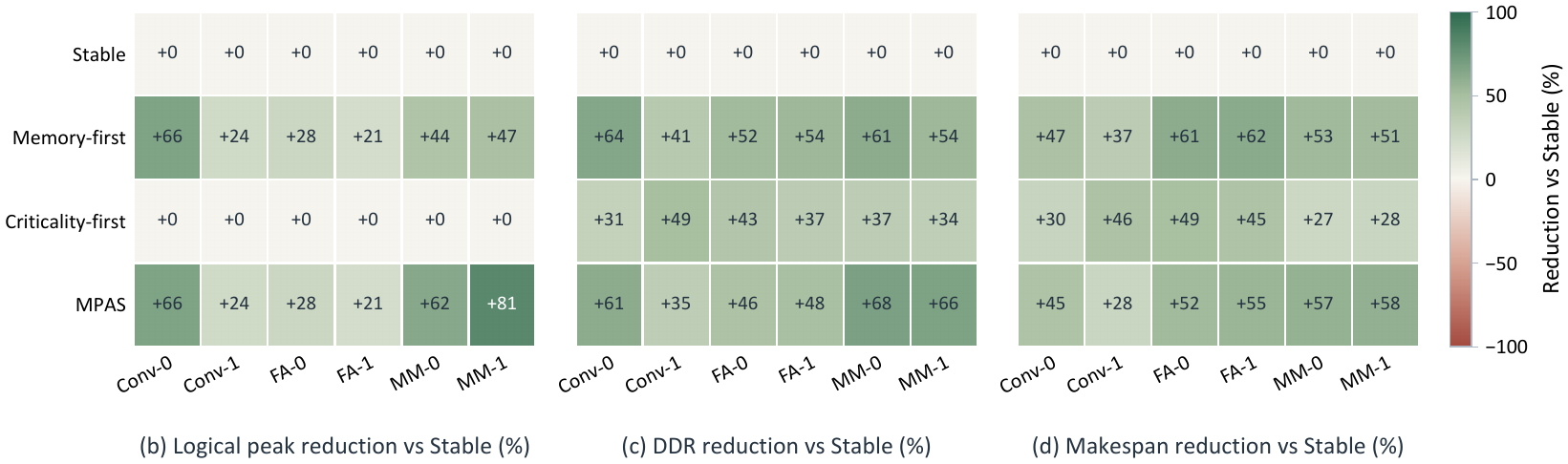}
    \vspace{-10pt}
  \caption{Stage and order analysis.
  Top: direct effect of each \lattice stage under the final configuration.
  Bottom: sensitivity to four legal orders using the same DLR/CPE backend;
  positive values denote reductions relative to Stable.}
    \vspace{-10pt}
  \label{fig:stage-order-analysis}
\end{figure*}

The top of Figure~\ref{fig:stage-order-analysis} isolates the contribution
of each stage along the directed compiler interface.  MPAS directly targets
the pre-placement order and reduces the geometric-mean logical peak by
\textbf{\MPASLogicalStageChange}.  With the MPAS order fixed, DLR's
cost-aware victim policy reduces DDR traffic by
\textbf{\DLRDDRStageChange} and spill-out count by
\textbf{\DLRSpillStageChange} relative to LRU\@.  With the complete DLR plan
fixed, CPE reduces geometric-mean makespan by
\textbf{\CPEGeomeanReduction}.

The per-workload results reflect the different structural bottlenecks shown
in Figure~\ref{fig:workload-characterization}.  MPAS produces its largest
logical-pressure reductions on Conv-0 and the matrix-multiplication traces,
where lifetime overlap is more pronounced.  CPE contributes most strongly on
the two FlashAttention traces, whose heterogeneous pipelines expose more
timing slack.  CPE leaves the layout, event stream, physical resident peak,
DDR traffic, and spill count unchanged on every workload, so its improvements
come exclusively from resource-order refinement.

The bottom of Figure~\ref{fig:stage-order-analysis} evaluates the four legal
orders under an otherwise identical backend.  Across the four orders, the
per-workload variation reaches \textbf{\OrderLogicalRange} in logical peak,
\textbf{\OrderPhysicalRange} in physical resident peak,
\textbf{\OrderDDRRange} in DDR traffic,
\textbf{\OrderSpillRange} in spill count, and
\textbf{\OrderMakespanRange} in makespan.  All 24 order--workload
combinations remain topologically legal and produce verified plans.

Criticality-first leaves logical peak unchanged but can reduce traffic and
makespan by advancing long dependency chains.  Memory-first and MPAS reduce
both lifetime pressure and downstream plan cost; MPAS further improves the
two matrix-multiplication logical peaks by incorporating criticality after
its lifecycle priorities.  The changing rankings across metrics confirm that
a legal order selects a distinct lifetime geometry and hence a distinct
physical memory and timing operating point.

\subsection{Memory-Plan Correctness and Timing Recovery}
\label{subsec:eval_contract}

\begin{figure*}[t]
  \centering
  \includegraphics[width=0.98\textwidth]
  {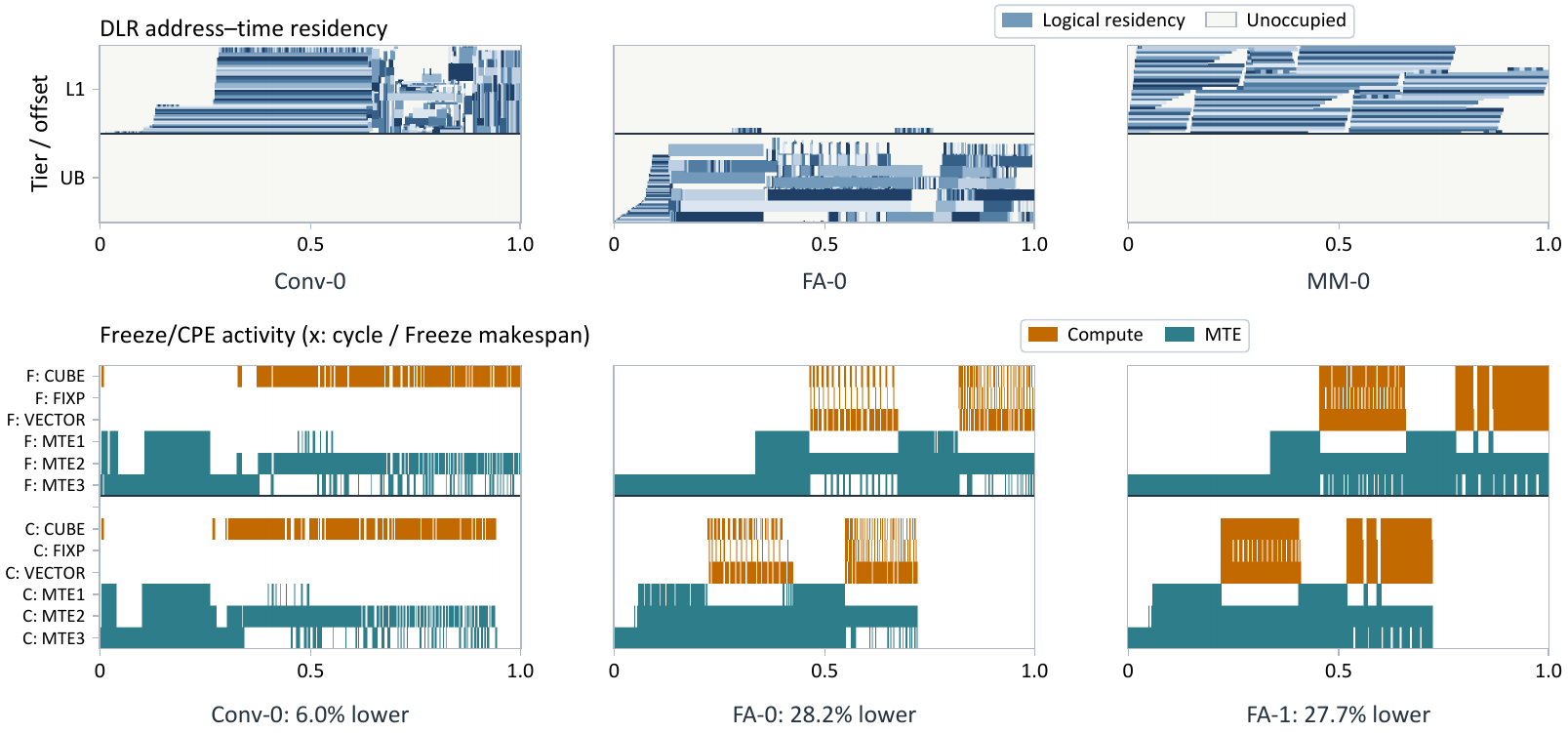}
  \caption{Materialized plans and timing refinement.
  Top: verified DLR address--time residency maps
  (blue: allocated residency; white: unoccupied space).
  Bottom: per-pipeline activity under Freeze and CPE on a common
  Freeze-normalized time axis (orange: compute; teal: MTE).}
    \vspace{-10pt}
  \label{fig:plan-timing-profiles}
\end{figure*}

\begin{table}[t]
\centering
\caption{Constraint-propagation diagnostics.
Ignore removes reuse constraints; Post-plan reorder changes command order
without replanning.  Makespan in (a) is normalized to Freeze.}
\label{tab:constraint_correctness}
\small
\textit{(a) Reuse enforcement on fixed MPAS+DLR plans.}\\[0.3ex]
\begin{tabular}{lrrr}
\toprule
\textbf{Policy} & \textbf{Valid} &
\textbf{Reuse violations} & \textbf{MS/Freeze} \\
\midrule
Ignore & 0/6 & 36,399 & 0.533 (infeasible) \\
Freeze & 6/6 & 0 & 1.000 \\
CPE & 6/6 & 0 & 0.879 \\
\bottomrule

\end{tabular}
\vspace{0.8ex}

\textit{(b) Phase-direction counterfactuals.}\\[0.3ex]
\begingroup
\setlength{\tabcolsep}{3.2pt}
\begin{tabular}{lrrr}
\toprule
\textbf{Sequence} & \textbf{Valid} &
\textbf{Static errors} & \textbf{Reuse violations} \\
\midrule
Post-plan reorder & 0/6 & 265{,}988 & -- \\
Timing without reuse & 0/6 & 0 & 36{,}399 \\
MPAS$\rightarrow$DLR$\rightarrow$CPE & 6/6 & 0 & 0 \\
\bottomrule

\end{tabular}
\endgroup
\end{table}

The top of Figure~\ref{fig:plan-timing-profiles} visualizes representative
DLR plans.  The horizontal axis follows normalized command order, while the
vertical ranges represent normalized UB and L1 address space.  Conv-0
develops dense late-stage reuse in L1; FA-0 uses shorter, more interleaved
residencies across UB; and MM-0 retains several long-lived L1 allocations.
Repeated use of the same address range at disjoint command intervals exposes
the ownership epochs from which DLR constructs $E_{\mathrm{reuse}}$.
Independent replay confirms that no address--time cell is simultaneously
owned by two residencies.

Table~\ref{tab:constraint_correctness}(a) tests whether these induced
constraints are required for valid execution.  \emph{Ignore} removes only
$E_{\mathrm{reuse}}$ while retaining input precedence, lifecycle, transfer,
residency, and resource constraints.  The verifier detects
\textbf{\IgnoreViolationTotal} ownership violations despite the apparent
timing reduction and rejects the resulting schedules for all
\textbf{\IgnoreInvalidWorkloads} workloads.
\emph{Freeze} preserves the
complete safe Stage-2 order, whereas CPE preserves $E_{\mathrm{fixed}}$ but
refines the remaining resource order.  CPE remains valid and strictly
improves all \textbf{\CPEImprovedWorkloads} workloads.

Table~\ref{tab:constraint_correctness}(b) verifies the direction of the
stage interfaces.  Reusing the event and address plan generated for the MPAS
order after replacing that order with Stable causes the verifier to identify
\textbf{\PostPlanStaticErrorTotal} event-anchor, residency, backing-state,
capacity, or address errors across
\textbf{\PostPlanInvalidWorkloads}/6 plans.  Optimizing timing before
materializing reuse constraints reproduces the
\textbf{\IgnoreViolationTotal} unconstrained ownership handoffs.  In contrast,
MPAS$\rightarrow$DLR$\rightarrow$CPE passes both replay layers on all six
workloads.  Once a memory plan has been materialized, later ordering changes
must therefore preserve its contract or trigger replanning.

The bottom of Figure~\ref{fig:plan-timing-profiles} illustrates how CPE
recovers valid overlap.  On Conv-0, FA-0, and FA-1, it reduces makespan by
6.0\%, 28.2\%, and 27.7\%, respectively, by changing cross-pipeline
alignment and local single-issue order.  The before/after replay confirms
identical placement, lifecycle and spill/reload events, physical resident
peak, DDR traffic, and spill count.  Thus, the shorter schedules capture the
timing freedom permitted by the fixed memory plan.

\subsection{Compiler Overhead}
\label{subsec:eval_compile}

\begin{figure}[t]
  \centering
  \includegraphics[width=\columnwidth]
  {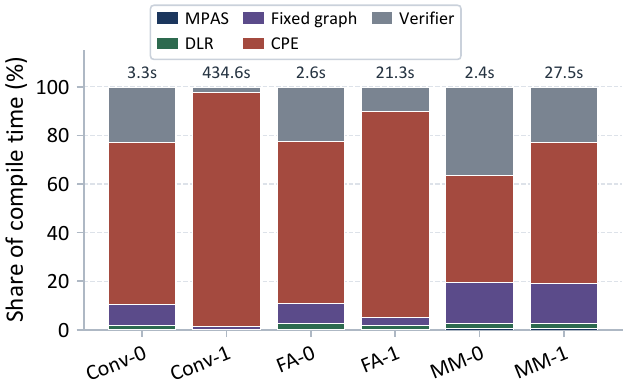}
  \vspace{-20pt}
  \caption{Compiler-time composition; totals appear above the bars.}
  \vspace{-10pt}
  \label{fig:compiler-overhead}
\end{figure}

Figure~\ref{fig:compiler-overhead} decomposes compilation into MPAS, DLR,
fixed-graph construction, CPE, and verification.  Total compile time ranges
from \textbf{\CompileTotalRange}.  MPAS and DLR remain lightweight; CPE
accounts for the largest share because it performs full timing evaluation and
verification for alternative resource orders.

This cost is incurred offline and can be amortized across repeated executions
of the generated plan.  CPE is also optional: disabling it returns the
verified MPAS+DLR+Freeze plan, while full \lattice applies one deterministic
seed and a bounded four-candidate local search.  This fixed optimization
envelope provides a direct search-quality--compile-time knob while preserving the
Stage-2 memory plan and its validity.

\section{Related Work}
\label{sec:related}

\textbf{Memory-aware scheduling and planning.}
Prior work reduces tensor residency through memory-aware graph ordering and
scheduling~\cite{ahn2020ordering,kumar2019efficient,levental2022memory,
wen2020timespace}, including hierarchical scheduling, capacity-aware block
grouping, energy-aware compilation, and pass orchestration
~\cite{wang2022hmcos,lee2023blockgroup,pao2025eda,yu2026opass}.
A complementary body of work addresses placement, spilling, offloading, and
rematerialization~\cite{jain2020checkmate,rhu2016vdnn,
wang2018superneurons,kirisame2021dtr,peng2020capuchin}, while Moccasin,
TelaMalloc, COSMA, and recent static planners coordinate allocation with
replacement or scheduling decisions~\cite{bartan2023moccasin,
maas2022telamalloc,li2023cosma,lamprakos2025futureproof,
steiner2023model}.  These studies establish the upstream coupling between
order and memory behavior.  \lattice targets the downstream boundary:
after a concrete layout and spill plan have been materialized, their
reuse-induced constraints become legality conditions for subsequent pipeline
refinement.

\textbf{ML/NPU compilation and execution.}
General ML compiler systems optimize graph lowering, tensor programs,
operator scheduling, and memory access
~\cite{chen2018tvm,roesch2018relay,zheng2020ansor,ma2020rammer,
zhu2022roller,shi2023welder,chen2024magis}.  Timeloop and VTA provide
architecture and mapping abstractions~\cite{parashar2019timeloop,
moreau2019vta}, while Tensor Comprehensions, TASO, PET, Optimus, and
learning-based optimization expand the transformation space
~\cite{vasilache2018tensorcomp,jia2019taso,wang2021pet,cai2022optimus,
chen2018learning}.  Accelerator-specific systems study PIM compilation, fused
dataflows, and NPU layout mapping
~\cite{sun2024pimcomp,symons2024stream,lee2025hopscotch}.
Inter-operator GPU and multi-NPU schedulers exploit coarser-grained
concurrency~\cite{guo2025interoperator,min2023flexer}, and NPU simulators,
compiler-space studies, platform benchmarks, and heterogeneous NPU--PIM
inference systems support broader architecture exploration
~\cite{hwang2023mnpusim,indirli2024npucompiler,jayanth2024edgebenchmark,
lai2024gem5nvdla,heo2024neupims}.  \lattice operates after operator lowering
on a static intra-core command DAG and is complementary to these higher-level
mapping, batching, and runtime decisions.

\textbf{Dependence-aware refinement.}
Compiler scheduling has long represented anti- and output dependencies
created by storage reuse, while integrated schedulers enforce memory,
register, and resource constraints
~\cite{desnos2013prepost,castaneda2019unison,lamprou2023safe}.
\lattice applies this principle to explicitly managed NPU memories: it derives
reuse dependencies from a tiered spill plan, propagates them across
MPAS/DLR/CPE, and distinguishes infeasible overlap from plan-preserving timing
refinement through the Ignore, Freeze, and CPE comparisons.

\section{Conclusion}
\label{sec:conclusion}

\lattice treats a static NPU memory plan as a first-class scheduling
contract.  \mpas shapes order-dependent lifetimes before address binding,
\dlr materializes placement, spill/reload behavior, and plan-induced reuse
constraints, and \cpe recovers pipeline parallelism while preserving the
selected plan.  Relative to the best evaluated external method for each
workload and metric, \lattice reduces peak memory, extra DDR traffic, spill
count, and modeled makespan by \textbf{\OverallPeakVsReference},
\textbf{\OverallDDRVsReference}, \textbf{\OverallSpillVsReference}, and
\textbf{\OverallMakespanVsReference}, respectively; plan-preserving \cpe
further reduces makespan by \textbf{\CPEGeomeanReduction} over Freeze while
preserving every memory metric.  The correctness diagnostics confirm that
the propagated memory constraints are necessary for executable timing.
More broadly, \lattice suggests that compilers for architectures with
explicitly managed memories and decoupled compute--transfer pipelines should
export memory-planning decisions as execution semantics, allowing subsequent
optimizations to proceed while retaining plan validity by construction.

\notice{\nocite{pei2025gcc,pei2026degs}}
\bibliographystyle{IEEEtranS}
\bibliography{refs}

\end{document}